\documentclass[fleqn,10pt]{wlscirep}
\usepackage[utf8]{inputenc}
\usepackage[T1]{fontenc}
\usepackage{subfigure}
\usepackage{booktabs}
\usepackage{tabularx,rotating}
\usepackage{multirow}
\usepackage{xr-hyper}
\usepackage{multirow}
\usepackage{colortbl}
\usepackage[nolists]{endfloat}

\title{Electricity supply quality and use among rural and peri-urban households and small firms in Nigeria}

\author[1,2]{Setu Pelz} 
\author[2,3]{Narges Chinichian} 
\author[2]{Clara Jütte} 
\author[2]{Philipp Blechinger}

\affil[1]{International Institute for Applied Systems Analysis, Laxenburg, Austria}
\affil[2]{Off-Grid Systems, Reiner Lemoine Institut (RLI), Berlin, Germany}
\affil[3]{Institute for Theoretical Physics, Technical University of Berlin, Germany}

\affil[*]{corresponding author(s): Philipp Blechinger (philipp.blechinger@rl-institut.de), Setu Pelz (pelz@iiasa.ac.at)}

\begin{abstract}
We present a household and enterprise energy survey dataset collected within the framework of the PeopleSuN project in Nigeria in 2021. Across three Nigerian geopolitical zones, a total of 3,599 households and 1,122 small and medium-sized enterprises were surveyed. The sample is designed to be representative of rural and peri-urban grid-electrified regions of each zone. Our surveys collect data on demographic and socioeconomic characteristics, energy access and supply quality, electrical appliance ownership and usage time, cooking solutions, energy related capabilities, and supply preferences. We encourage academic use of the data presented and suggest three avenues of further research: (1) modelling appliance ownership likelihoods, electricity consumption levels and energy service needs in un-electrified regions; (2) identifying supply-side and demand-side solutions to address high usage of diesel generators; (3) exploring broader issues of multi-dimensional energy access, access to decent living standards and climate vulnerability.
\end{abstract}
\begin{document}

\flushbottom
\maketitle

\thispagestyle{empty}

\section*{Background \& Summary}

Rural and peri-urban populations in Nigeria continue to suffer unreliable and expensive energy supply. According to the World Bank, the electricity access rate in Nigeria stood at 55.4\% in 2020 with a big gap between urban and rural areas (83.9\% vs. 24.6\%)\cite{WorldBank2020}. At the same time, nearly 30 million Nigerian households depend on wood as a source of fuel, the collection of which is time consuming and mainly done by women\cite{world2022world}. Where there is supply, it is typically unreliable and frequently interrupted by blackouts. The Nigeria Enterprise Survey from the World Bank showed that 27\% of Nigerian firms identified reliability of electricity supply as the main obstacle to their business\cite{WBES2014}. On average, 32.8 power outages were reported to occur in a typical month leading to an estimated 11\% loss in sales value\cite{WBES2014}. The average grid-connected household receives just 6.6 hours of supply on a typical day, linked to a per capita consumption of just 144kWh per year\cite{LSMS2019}. In comparison, the annual per capita consumption in Ghana and South Africa is respectively 351 kWh and 4,198 kWh. Plagued by issues of supply quality, many Nigerians have resorted to self-generation using petrol and diesel generators, spending approximately 1.56 trillion Naira (3.76 billion USD\footnote{Average exchange rate in 2021: 1 Naira = 0.0024 USD}) per year on fuel \cite{Ugwoke2020}. 

A central motivation for improving energy supply is to provide the means to secure access to basic energy service needs \cite{fell_energy_2017}. These needs include examples such as `being at a comfortable temperature' or `being able to store food'. These are closely linked with the broader concept of `basic human needs' \cite{doyal_theory_1991} and have been used to define minimum material (energy) requirements under the Decent Living Standards \cite{rao_decent_2018}. Data describing satisfaction of such needs is severely limited to non-existent, and as we show through this data release, cannot be conflated with mere access to a modern energy carrier. An analysis of household access to modern energy carriers and supply quality must therefore be complemented by an analysis of satisfaction of a set of basic energy services. Indeed this notion of energy service needs is a foundational element of the World Bank's multi-tier framework for measuring energy access \cite{Bhatia2015, ESMAP2020}. Moreover, an understanding of energy service needs and preferences is especially important for planning off-grid electrification measures where appliances and energy carriers are typically bundled, forming energy service delivery systems. Several studies and analyses have emphasized that in order for Nigeria to reach its access targets while also increasing the proportion of renewable energy in its energy mix, a combination of grid extension and off-grid systems, e.g. mini-grids, and solar home systems (SHS) is necessary \cite{IEA2019, esmap2022, SE4All2022}. 

While global efforts are accelerating under the banner of achieving Sustainable Development Goal 7 (SDG7) by 2030, stakeholder discussions indicate that progress in Nigeria is hindered by data availability, among other barriers. Data describing the energy access deficit in Nigeria exists (see Table \ref{tab:otherdata}), however, there is limited disaggregate information describing the supply quality in the existing network and the unmet demand in `un-electrified' regions. In this paper we present survey data collected to fill this and other gaps through the `People Power: Optimizing off-grid electricity supply systems in Nigeria' project (PeopleSuN)\footnote{PeopleSuN is funded by the German Federal Ministry of Education and Research (BMBF) within the funding initiative `Client II - International Partnerships for Sustainable Innovations'.}. The methods applied draw on stakeholder discussions conducted in Nigeria to define the data gap and the necessary survey and sampling strategy to address this. The questionnaires used draw from specific modules within established surveys capturing energy-related data, most directly from the Multi-Tier Framework for Measuring Energy Access surveys \cite{Bhatia2015}. The surveys provide data on household and enterprise characteristics, energy supply and consumption. They also capture preferences, trust in institutions and several gender-disaggregated variables. Our sample is representative of grid-electrified rural and peri-urban regions across three geopolitical zones with large energy access deficits. This data provides important insight into actual energy supply quality and use among `electrified' communities and can be used to improve models of energy demand in similar but currently `un-electrified' communities. The final sample includes 3,599 households and 1,122 small and medium sized enterprises from 225 enumeration areas across the three geopolitical zones.

\begin{table}[ht!]
\centering
\scalebox{0.8}{
\begin{tabular}{p{0.2\linewidth}p{0.15\linewidth}p{0.15\linewidth}p{0.15\linewidth}p{0.15\linewidth}}
\hline
\textbf{Data source} & \textbf{Description} & \textbf{Year} & \textbf{Representativeness} & \textbf{Energy data} \\
\hline
General Household Survey / Living Standards Measurement Study (World Bank) \url{https://www.worldbank.org/en/programs/lsms} & National household living standards survey. & Multiple survey rounds (most recent 2018) & Nationally representative survey of approximately 5,000 households, which are also representative of the six geopolitical zones. & Household socioeconomics, assets and general energy access questions. \\
\hline
Nigeria SE4ALL Platform. Federal Ministry of Power, Nigeria. \url{https://nigeriase4all.gov.ng/} & Platform provides access to geospatial data and market data related to grid-connected and off-grid energy in Nigeria. & Frequently updated. & Geospatial data for all of Nigeria. & Spatial data describing locations of consumers, existing and planned grid and off-grid energy infrastructure, and administrative boundaries. \\
\hline
Energy Sector Management Assistance Program (ESMAP), World Bank. Nigeria - Multi-Tier Framework for Measuring Energy Access Household Survey (MTF) 2018. \url{https://microdata.worldbank.org/index.php/catalog/3865} & The MTF survey assesses energy access and use in North, North-West Nigeria. & 2018 & Stratified random sampling, with equal allocation between urban and rural areas and equal allocation between grid-user and non grid-user households in North, North-West of Nigeria. & Data on energy access, types of energy sources used, appliance ownership and energy expenditure. \\
\hline
Falchetta, G., Pachauri, S., Parkinson, S. et al. A high-resolution gridded dataset to assess electrification in sub-Saharan Africa. Sci Data 6, 110 (2019). \url{https://doi.org/10.1038/s41597-019-0122-6} & A high-resolution gridded dataset to assess electrification in sub-Saharan Africa & 2019 & sub-Saharan Africa & Spatial distribution and temporal evolution of electricity access. \\
\hline
DHS surveys - Nigeria \url{https://dhsprogram.com/} & Demographic and Health Surveys collect data on population, health, and nutrition in Nigeria. & Multiple survey rounds (most recent 2018) & Nationally representative, with data available for individual states and regions. & Data on household electricity access, types of cooking fuels, and ownership of appliances. \\
\hline
World Bank Enterprise Surveys for Nigeria \url{https://www.enterprisesurveys.org/} & Enterprise surveys provide data on the business environment, including energy-related aspects, through interviews with firm managers and owners. & Periodically conducted, most recent in 2014. & Covers various regions in Nigeria with a focus on the formal sector. & Energy reliability, power outages, and energy expenditures for businesses. \\
\hline
\end{tabular}}
\caption{\textbf{Overview of existing sources of data describing energy access and use in rural and peri-urban Nigeria.}}
\label{tab:otherdata}
\end{table}

\section*{Methods}

\begin{figure}[ht!]
\centering
\includegraphics[width=0.9\textwidth]{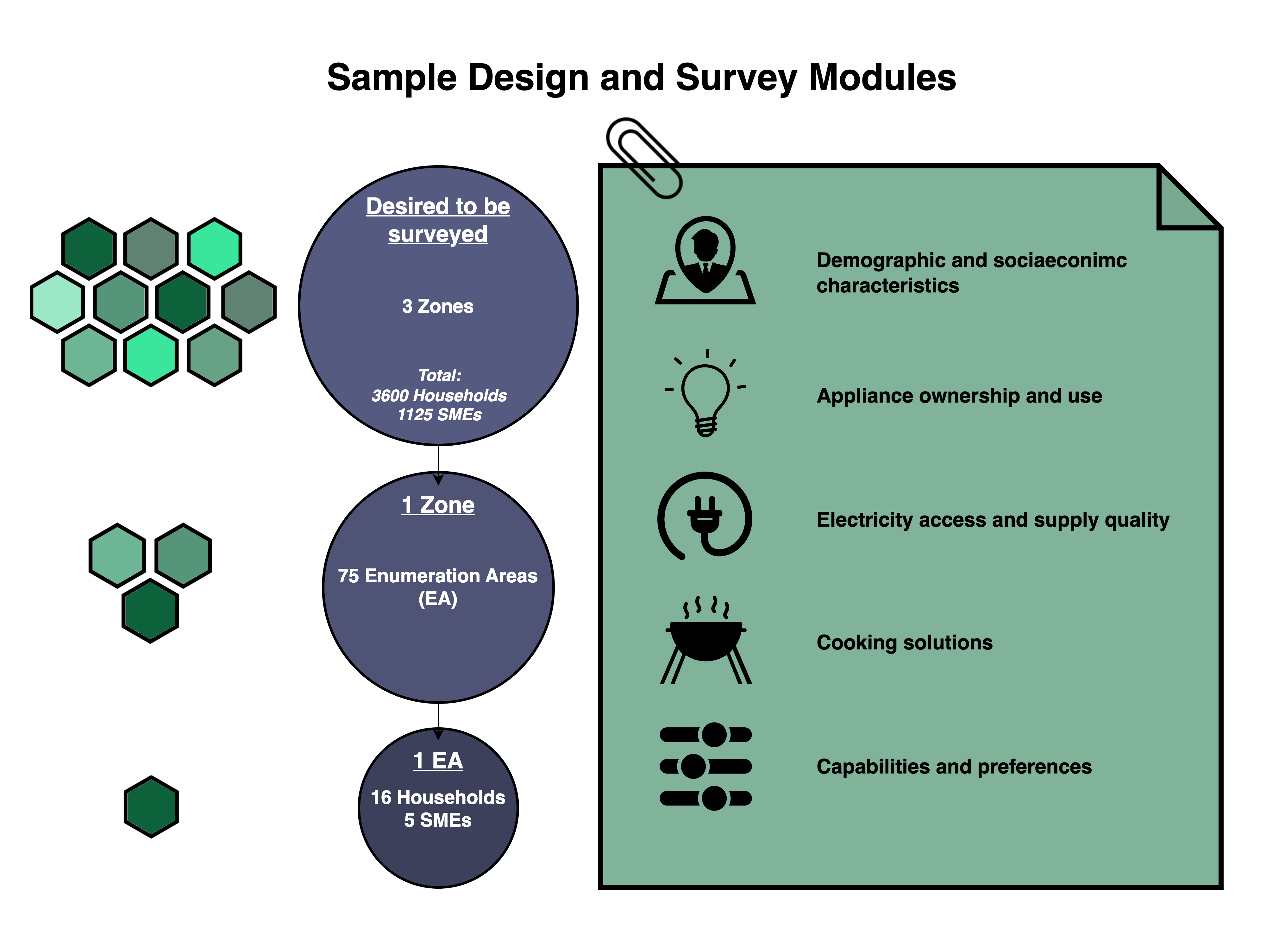}
\caption{\textbf{Sample design and questionnaire modules}. Three of six Nigerian zones, North West, North Central and South South were selected. These zones were chosen based on lower urbanisation rates (excluding the South West and South East) and safety and logistical considerations (excluding North East). Seventy five enumeration areas (EA) were marked in each zone [total of 225 EAs]. Within each enumeration area, sixteen households and five enterprises were planned to be surveyed. The desired distribution of EAs and the final implementation are shown respectively in Table \ref{tab:sampleoriginal}. The surveys cover a wide range of questions regarding demographic and socioeconomic variables, electricity availability and supply quality, appliance ownership and use, available cooking solutions,  capabilities and preferences.}
\label{fig:schematic}
\end{figure}

Figure \ref{fig:schematic} provides an overview of the sample design and survey modules used. Complete codebooks describing the questionnaires are available here: \url{
https://dataverse.harvard.edu/dataset.xhtml?persistentId=doi:10.7910/DVN/GTNEJD}. We now describe the data collection methods in detail.

\subsection*{Sample frame}
\label{sect:sample}

\begin{figure}[ht!]
\centering
\includegraphics[width=0.8\textwidth]{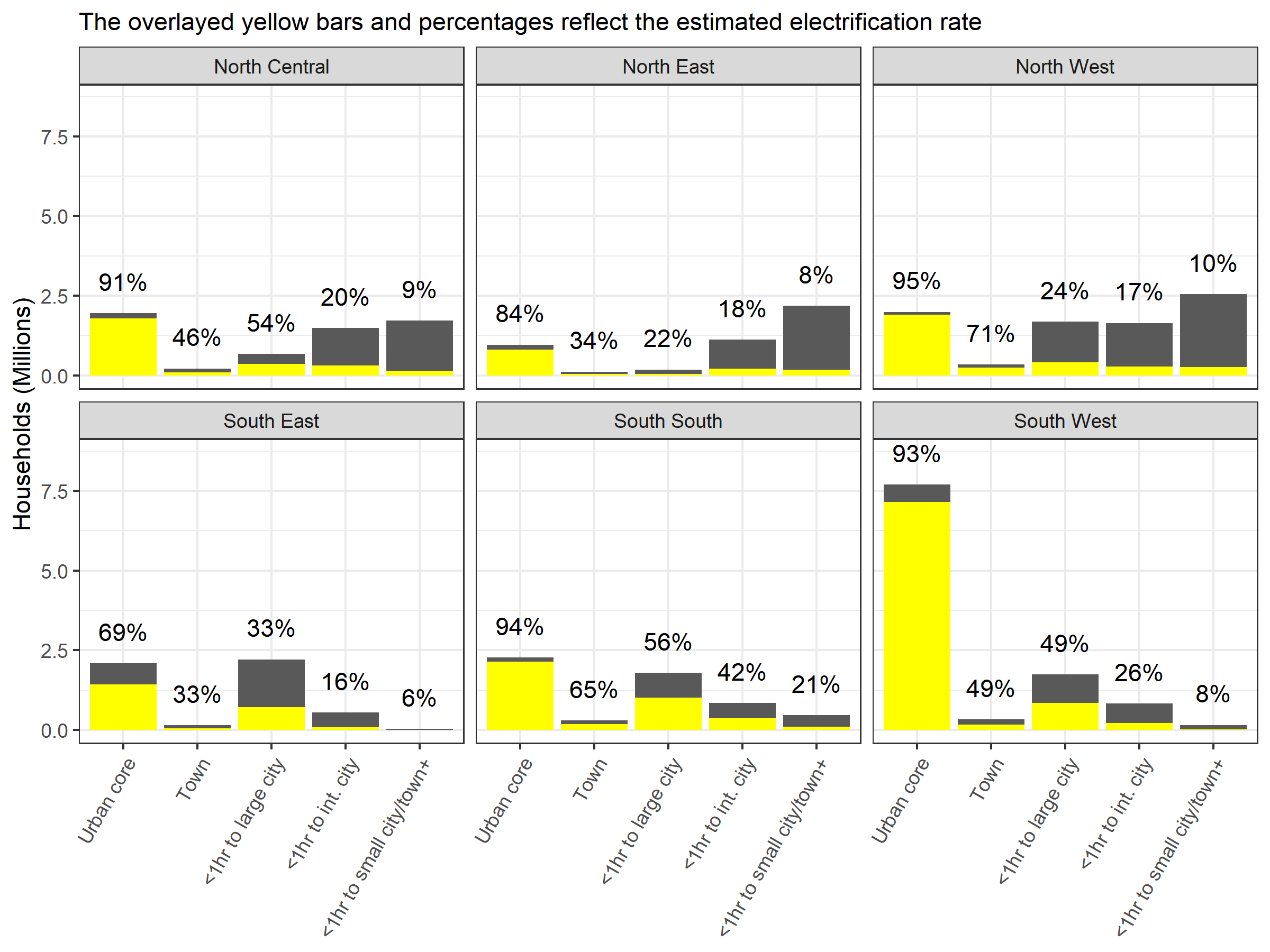}
\caption{\textbf{Distribution of households and share electrified by rurality across Nigeria}. Population estimates are taken from the latest bottom-up WorldPop dataset for Nigeria (V1.2, see \cite{Bondarenko2020}). These were converted to household estimates using the average household size in each state as captured by the 2018 DHS survey (see \cite{dhs2018}). Rurality is defined using an aggregation of the URCA definition (see \cite{Cattaneo2021}). Each group reflects a more rural region, starting with cities of all sizes which are aggregated to the group `Urban core', to very remote areas including those further than 1 hour from a small city or town which are aggregated into the final group `<1hr to small city/town+'. Electrification estimates are taken from AtlasAI spatial electrification data for the year 2020 (see atlasai.com). Our work focusses on grid-electrified households outside the urban core.}
\label{fig:household_distribution}
\end{figure}

Three geopolitical zones were selected for sampling: North West, North Central, and South South. These zones were purposefully chosen based on lower rates of urbanization (excluding South West and South East) and safety and associated logistical concerns (excluding North East). The sample frame was then developed using remotely sensed datasets due to the lack of up-to-date administrative data. Four remotely sensed spatial datasets were used, as shown in Table \ref{tab:sampleframe}. The focus of this work was on understanding energy supply quality and use among grid-electrified households living in rural and peri-urban regions (outside urban centers). Urbanization was defined using the `Urban Rural Catchment Area' dataset (URCA), which stratified the population by travel time to the nearest agglomeration. Further detail on the URCA sub-categories and their definition can be found in \cite{Cattaneo2021}. For the purposes of this work, the URCA categories were aggregated into an Urban-core and varying degrees of rurality, as shown in Table \ref{tab:urca}. Figure \ref{fig:household_distribution} describes the distribution of households across these categories and separated into those assumed to be electrified as per the ELEC dataset. All Towns (>0.02M and <0.05M population) and regions that were within or further than 1 hour travel time from a city boundary were defined as outside the urban core and relevant to this study. Estimated population sizes and electrification rates within each aggregate spatial category across all six geopolitical zones are shown in Figure \ref{fig:household_distribution}, providing further intuition regarding the selection of North Central, North West, and South South as zones of study.

\begin{table}[ht!]
\centering
\scalebox{.8}{
\begin{tabular}{p{0.1\linewidth}p{0.4\linewidth}p{0.2\linewidth}p{0.25\linewidth}}
\hline \hline
\textbf{Acronym} & \textbf{Detail} & \textbf{Resolution / Year} & \textbf{Source} \\ \hline
(WPOP) & WorldPop Bottom-up Population Estimate & 0.1km$^2$, 2020 & \cite{Bondarenko2020} \\ [10mm]
(URCA) & Urban-Rural Catchment Area & 1km$^2$, 2015 & \cite{Cattaneo2021} \\ [10mm]
(ELEC) & AtlasAI Electrification Data & 4km$^2$, 2020 & (see atlasai.com) \\ [10mm]
(ASSET) & AtlasAI Asset Index & 4km$^2$, 2020 & (see atlasai.com) \\ \hline
\end{tabular}}
\caption{\textbf{Overview of remotely sensed datasets used in defining the sample frame.}}
\label{tab:sampleframe}
\end{table}

\begin{table}[ht!]
\centering
\subtable[URCA original categories]{
\scalebox{.8}{
\begin{tabular}{rrl|ccccccc}
\multicolumn{3}{c|}{\multirow{2}{*}{\textbf{}}}                               & \multicolumn{2}{c}{Large cities} & \multicolumn{2}{c}{Intermediate cities} & \multicolumn{2}{c}{Small cities} & Towns      \\
\multicolumn{3}{c|}{}                                                         & \textgreater{}5M      & 1-5M     & 0.5-1M            & 0.25-0.5M           & 0.1-0.25M       & 0.05-0.1M      & 0.02-0.05M \\ \hline
\multicolumn{3}{r|}{Urban centers}                                            & 1                     & 2        & 3                 & 4                   & 5               & 6              & 7          \\
Peri-urban                       & \multicolumn{2}{r|}{0-1 hours}             & 8                     & 9        & 10                & 11                  & 12              & 13             & 20a        \\
\multirow{2}{*}{Peri-rural}      & \multicolumn{2}{r|}{1-2 hours}             & 14                    & 15       & 16                & 17                  & 18              & 19             & 20b        \\
                                 & \multicolumn{2}{r|}{2-3 hours}             & 21                    & 22       & 23                & 24                  & 25              & 26             & 27         \\
Hinterlands                      & \multicolumn{2}{r|}{\textgreater{}3 hours} &                       &          &                   &                     &                 &                & 28/29     
\end{tabular}}}
\subtable[URCA PeopleSuN aggregated categories]{
\scalebox{.8}{
\begin{tabular}{rrl|ccccccc}
\multicolumn{3}{c|}{}                                                          & \multicolumn{2}{c}{Large cities}                & \multicolumn{2}{c}{Intermediate cities}           & \multicolumn{2}{c}{Small cities} & Towns                         \\
\multicolumn{3}{c|}{\multirow{-2}{*}{\textbf{}}}                               & \textgreater{}5M              & 1-5M            & 0.5-1M                 & 0.25-0.5M                & 0.1-0.25M       & 0.05-0.1M      & 0.02-0.05M                    \\ \hline
\multicolumn{3}{r|}{Urban centers}                                             & \multicolumn{6}{c}{\cellcolor[HTML]{757171}Urban Core}                                                                                 & \cellcolor[HTML]{FFF2CC}Cat 1 \\
Peri-urban                        & \multicolumn{2}{r|}{0-1 hours}             & \multicolumn{2}{c}{\cellcolor[HTML]{FFE699}Cat 2} & \multicolumn{2}{c}{\cellcolor[HTML]{FFD966}Cat 3} & \multicolumn{3}{c}{\cellcolor[HTML]{BF9000}Cat 4}                \\
                                  & \multicolumn{2}{r|}{1-2 hours}             & \multicolumn{7}{c}{\cellcolor[HTML]{BF9000}Cat 4}                                                                                                                      \\
\multirow{-2}{*}{Peri-rural}      & \multicolumn{2}{r|}{2-3 hours}             & \multicolumn{7}{c}{\cellcolor[HTML]{BF9000}Cat 4}                                                                                                                      \\
Hinterlands                       & \multicolumn{2}{r|}{\textgreater{}3 hours} &                               &                 &                        &                          &                 &                & \cellcolor[HTML]{BF9000}Cat 4
\end{tabular}}}
\caption{\textbf{Original URCA spatial categories and corresponding PeopleSuN aggregated categories.} URCA categories reflect distinct catchment areas describing distance from urban agglomerations taking into account road quality among other factors. These are aggregated for the purposes of the PeopleSuN sample. See \cite{Cattaneo2021} for further detail regarding the categorisation approach.}
\label{tab:urca}
\end{table}

\subsection*{Sampling}

In total 3,600 households were intended to be surveyed, split evenly into the three geopolitical zones. Households were sampled from 225 enumeration areas (EAs) consisting of 16 households each. In addition, 5 firms were to be surveyed in each enumeration area, giving a total sample of 1,125 firms. Within each zone, half of the states (four in total per state) were selected using systematic probability proportionate to size sampling (PPS) in order to reduce survey costs. The count of households at the state level was used as the weighting variable in the PPS sampling. This approximates a representative zonal sample when not able to visit all states within each zone. The state-level pre-sampling may introduce bias with respect to the rurality of sparsely populated states not included in our sample. To address this the distribution of households with respect to the URCA PeopleSuN categories across \textit{all} states in each zone is used to adjust subsequent sampling within selected states. Table \ref{tab:sampleoriginal} describes the approximate proportion of grid-connected households in each zone and category. This is done including all states, not just those sampled. The proportions therefore represent the share of people living in each category that should be replicated within the restricted sample of states. This proportional stratification results in a sample size within each zone and category that reflects, as far as possible, a representative sample of grid-electrified rural and peri-urban households across the zone.

The final enumeration areas within each zone are selected using PPS sampling once again, weighted by the total number of households in each cell to achieve the sample size by zone as defined in the Desired EAs column. A total sample of 225 enumeration areas is equivalent to a zonal sample of 75 enumeration areas, distributed as per the proportional stratification across each of the three zones. Within each zone, this results in a sample that is representative of grid-electrified households outside the urban core while reducing the cost of survey implementation. Analysis can therefore be conducted at the zonal level as is, and re-weighted to arrive at a representative `national' sample across all three contiguous zones if desired. The necessary weights for aggregate analyses across all three contiguous zones are provided with the dataset.

\subsection*{Implementation}

Survey implementation was conducted by eHealth Africa (eHA), who also secured state-government permission to conduct surveys in each state. The survey was conducted using the Kobo Collect Computer-Assisted Personal Interviews (CAPI) technology. All EAs within security compromised Local Government Areas (LGAs) were identified and shared with the research team for replacement. From the total of 255 EAs originally planned (reflecting a desired sample of 225 EAs and 30 buffer EAs), 247 EAs remained during implementation. From the 21st to the 23rd June 2021, a training of trainers was conducted by the research team for eHA’s key staff and field supervisors who were scheduled to work on the project. Following this, the translation of survey instruments into local languages (Hausa and Pidgin) was finalized on the 12th of July. This was followed by field testing which was conducted on the 13th and 14th of July 2021, and a field testing report was submitted on 16th July 2021. Step-down training of Enumerators was simultaneously conducted between 25th - 29th July 2021 in 3 locations (Kano, Abuja, and Akwa Ibom) to accommodate the enumerators by zone and reduce travel costs. Kano hosted the North West states’ enumerators while Abuja and Akwa-Ibom hosted North Central and South-South enumerators respectively with the report submitted on the 30th of July 2021. Pre- and post-training assessments were conducted and enumerators that met the evaluation benchmarks were finally selected. A total of 60 Enumerators and 12 State Supervisors were engaged in the quantitative survey activity. 

\begin{figure}[ht!]
\centering
\includegraphics[width=0.9\textwidth]{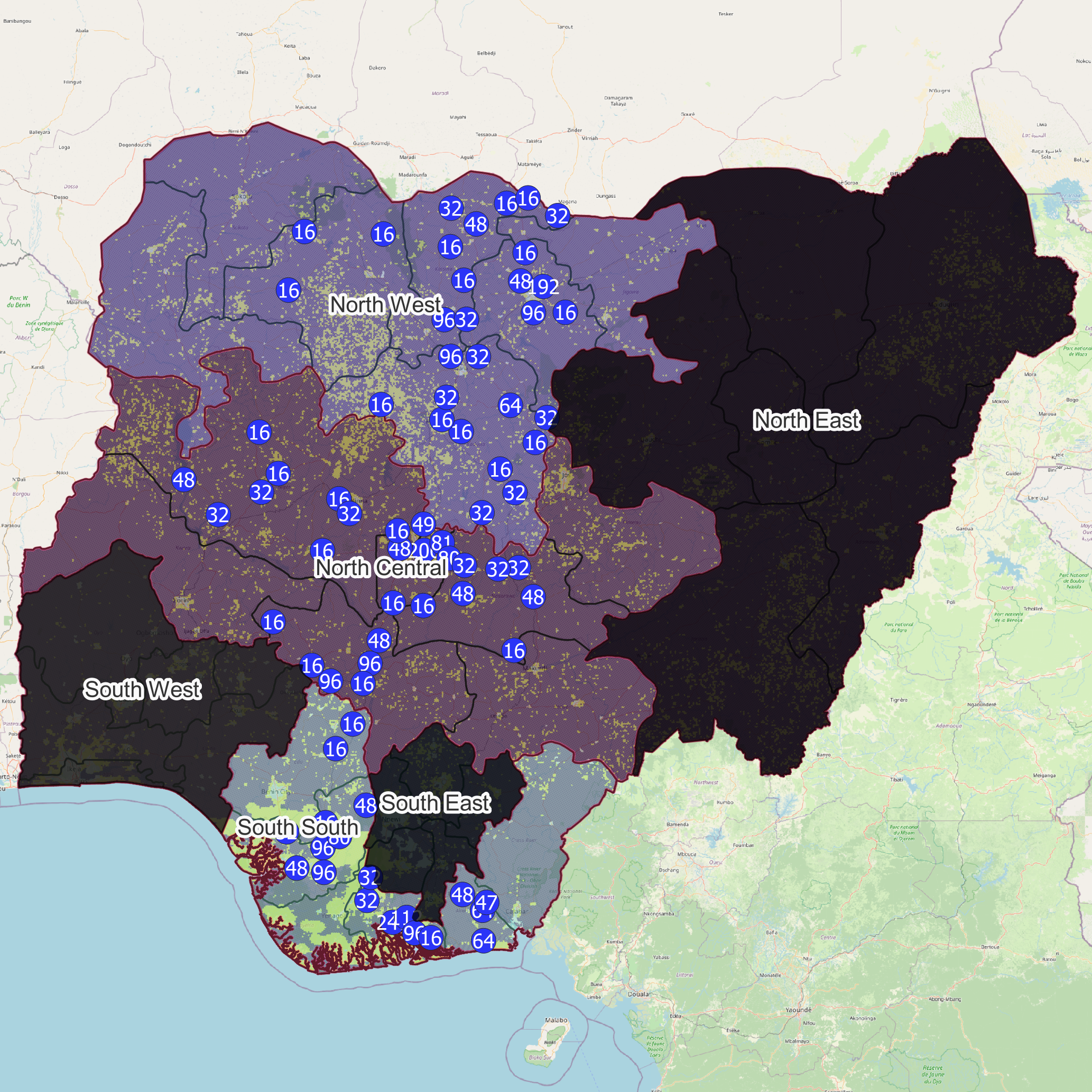}
\caption{\textbf{Visualisation of the spatial distribution of the household sample}. Geopolitical zones that were not sampled are shaded dark grey. Actual sample locations have been jittered up to 10 kilometres and labels reflect the total number of households sampled (numbers greater than 16 indicate a cluster of enumeration areas). The basemap includes the spatial electrification coverage for 2020 as estimated by AtlasAI, shown as yellow pixels. Any geographic boundary inaccuracies are unintentional. All secondary data is from identical sources as stated for Figure \ref{fig:household_distribution}.}
\label{fig:fullimage_sample}
\end{figure}

Data collection commenced on August 7th. A team comprising 2 enumerators (mostly paired male and female) worked to cover 16 households and 5 enterprises in each EA. Community leaders and local authorities were consulted before enumerators’ visits and commencement of activity at the LGA and community levels. The team faced challenges in gathering survey data due to security, communal clashes and inaccessibility in certain EAs, requiring replacement with a list of buffer enumerations areas within each zone. As a result, there is some small level of bias with respect to the zones given the mismatch in URCA group proportions, which is shown in Table \ref{tab:sampleoriginal}. This remains small overall and the mechanism underlying this transparently communicated here. Furthermore, 1 enumeration area in North Central, and 6 in North West were not grid-connected (despite being identified as electrified within the AtlasAI data), but were using electricity in other forms such as diesel generators or solar devices. One EA (AK12) contains a sample of 15, rather than 16 households which was not able to be completed due to banditry on the final day of enumeration, giving us a final sample of 3,599 households. Figure \ref{fig:fullimage_sample} depicts the full sample of enumeration areas with household counts shown as labels. Points have been jittered up to 10km. 

\begin{table}[ht!]
\centering
\scalebox{.8}{
\begin{tabular}{llrrrrr}
\hline \hline
\textbf{Zone} & \textbf{URCA PeopleSuN category}         & \textbf{Households} & \textbf{Share} & \textbf{Desired EAs} & \textbf{Sampled EAs} & \textbf{Grid-connected} \\ \hline
North Central & Town (0.02M)            & 98297               & 0.11           & 8   & 9 & 9                \\
North Central & 1hr to large city       & 364067              & 0.4            & 30 & 30 & 30                   \\
North Central & 1hr to int. city        & 301774              & 0.33           & 25   & 25 & 24                \\
North Central & 1hr to small city/town+ & 147266              & 0.16           & 12   & 11 & 11                \\
North West    & Town (0.02M)            & 243730              & 0.2            & 15  & 17 & 17                 \\
North West    & 1hr to large city       & 413041              & 0.35           & 26 & 27 & 26                  \\
North West    & 1hr to int. city        & 275368              & 0.23           & 17  & 18 & 18                 \\
North West    & 1hr to small city/town+ & 261740              & 0.22           & 17   & 13 & 8                \\
South South   & Town (0.02M)            & 188965              & 0.11           & 8 & 10 & 10                   \\
South South   & 1hr to large city       & 1012942             & 0.61           & 46  & 46 & 46                 \\
South South   & 1hr to int. city        & 356610              & 0.22           & 17   & 16 & 16                \\
South South   & 1hr to small city/town+ & 98675               & 0.06           & 5   & 3 & 3         \\ \hline        
\end{tabular}}
\caption{\textbf{Desired and final enumeration areas per zone and URCA PeopleSuN category}. Households describes the approximate number of grid-connected households in each category. Desired EAs describes the desired allocation of enumeration areas providing a representative sample of rural and peri-urban grid-connected households in each zone. Sampled EAs describes the actual final sample following implementation and Grid-connected describes those that were indeed grid-connected.}
\label{tab:sampleoriginal}
\end{table}

\section*{Data Records}

All data is publicly available here \url{
https://dataverse.harvard.edu/dataset.xhtml?persistentId=doi:10.7910/DVN/GTNEJD}

\section*{Summary statistics}

As this is a novel survey dataset with a transparent sampling design described above, we provide here an overview of the collected data. This includes sample characteristics, lighting sources used by households and enterprises, self-reported supply quality, appliance ownership, cooking solutions, and energy service satisfaction. Several additional variables exist in the dataset beyond that which are shown here.

\subsection*{Sample characteristics}

Summary statistics of the household and enterprise samples are shown in Table \ref{tab:samplesummary} panels (a) and (b). These aggregates provide an indication of sampled household and enterprise characteristics across common socio-economic, demographic and asset-wealth variables.

\begin{table}[ht!]
\centering
\subtable[Household]{
\scalebox{.6}{

\centering
\begin{tabular}[t]{lrrr}
\toprule
Variable & North Central & North West & South South\\
\midrule
Age of household head & 47.25 & 47.39 & 45.80\\
Household adults & 4.19 & 4.56 & 3.21\\
Number of bicycles & 0.14 & 0.43 & 0.23\\
Number of motorbikes & 0.62 & 0.61 & 0.42\\
Number of cars & 0.37 & 0.16 & 0.32\\
Share using Bank & 0.79 & 0.54 & 0.86\\
Share using Mobile Money & 0.24 & 0.04 & 0.27\\
Weekly (non-energy) expenditures & 10317.08 & 12643.75 & 13758.67\\
\bottomrule
\end{tabular}

}}
\subtable[Enterprise]{
\scalebox{.6}{

\centering
\begin{tabular}[t]{lrrr}
\toprule
Variable & North Central & North West & South South\\
\midrule
Age of owner & 37.21 & 34.53 & 39.11\\
Female owner & 0.66 & 0.09 & 0.59\\
Number of bicycles & 0.02 & 0.22 & 0.09\\
Number of motorbikes & 0.37 & 0.48 & 0.34\\
Number of cars & 0.15 & 0.06 & 0.15\\
Share using Bank & 0.62 & 0.45 & 0.71\\
Share using Mobile Money & 0.18 & 0.05 & 0.27\\
Total assets & 212658.71 & 207399.19 & 271132.80\\
Full-time employees & 1.62 & 1.38 & 1.52\\
Monthly salaries & 23735.14 & 26444.35 & 24694.40\\
\bottomrule
\end{tabular}

}}
\caption{\textbf{Sample summary statistics across common socioeconomic, demographic and asset-wealth characteristic variables.} All monetary aggregates are provided in Nigerian Naira (2021).}
\label{tab:samplesummary}
\end{table}

Table \ref{tab:access_commhh} provides sample aggregates of the rates of grid electrification at the community-level, together with the household and enterprise access to electricity. Notably, we find that over half of the households interviewed also owned backup sources ranging from electric generators to rechargeable and dry-cell batteries. We also note once more that despite efforts to establish a sample of grid-electrified communities, our sample indeed includes a small subset of communities that were not grid-electrified, but utilize other sources of electricity, as described in the sampling design and implementation section above.

\begin{table}[ht!]
\centering
\scalebox{.67}{

\centering
\begin{tabular}[t]{lrrr}
\toprule
Access & North Central & North West & South South\\
\midrule
\addlinespace[0.3em]
\multicolumn{4}{l}{\textbf{Community-level}}\\
\hspace{1em}National grid connection & 99\% & 92\% & 97\%\\
\addlinespace[0.3em]
\multicolumn{4}{l}{\textbf{Household-level}}\\
\hspace{1em}National grid connection & 98\% & 85\% & 93\%\\
\hspace{1em}Local minigrid & 0\% & 0\% & 4\%\\
\hspace{1em}Solar home system & 1\% & 6\% & 1\%\\
\hspace{1em}Solar lantern & 1\% & 2\% & 0\%\\
\hspace{1em}Electric generator & 37\% & 18\% & 66\%\\
\hspace{1em}Rechargeable battery & 35\% & 26\% & 30\%\\
\hspace{1em}Dry cell battery / torch & 59\% & 68\% & 32\%\\
\hspace{1em}Other & 10\% & 5\% & 8\%\\
\bottomrule
\end{tabular}

}
\caption{\textbf{Aggregate community-, household-, and enterprise-level energy access rates.} Top: Share of enumeration areas that were in-fact grid connected (note: as described in the research design, the intended sample of grid-electrified communities included a minority of communities that were not grid-connected by consumed electricity in some way). Bottom: Share of households and enterprises using each possible source of lighting. The latter adds to more than 100\% due to fuel stacking discussed later. See Table \ref{tab:prisecelec} for further detail describing patterns of stacking of each of the above solutions for the households and enterprises. }
\label{tab:access_commhh}
\end{table}

\subsection*{Household and enterprise electricity supply stacking}

\begin{figure}[ht!]
\centering
\includegraphics[scale = 0.6]{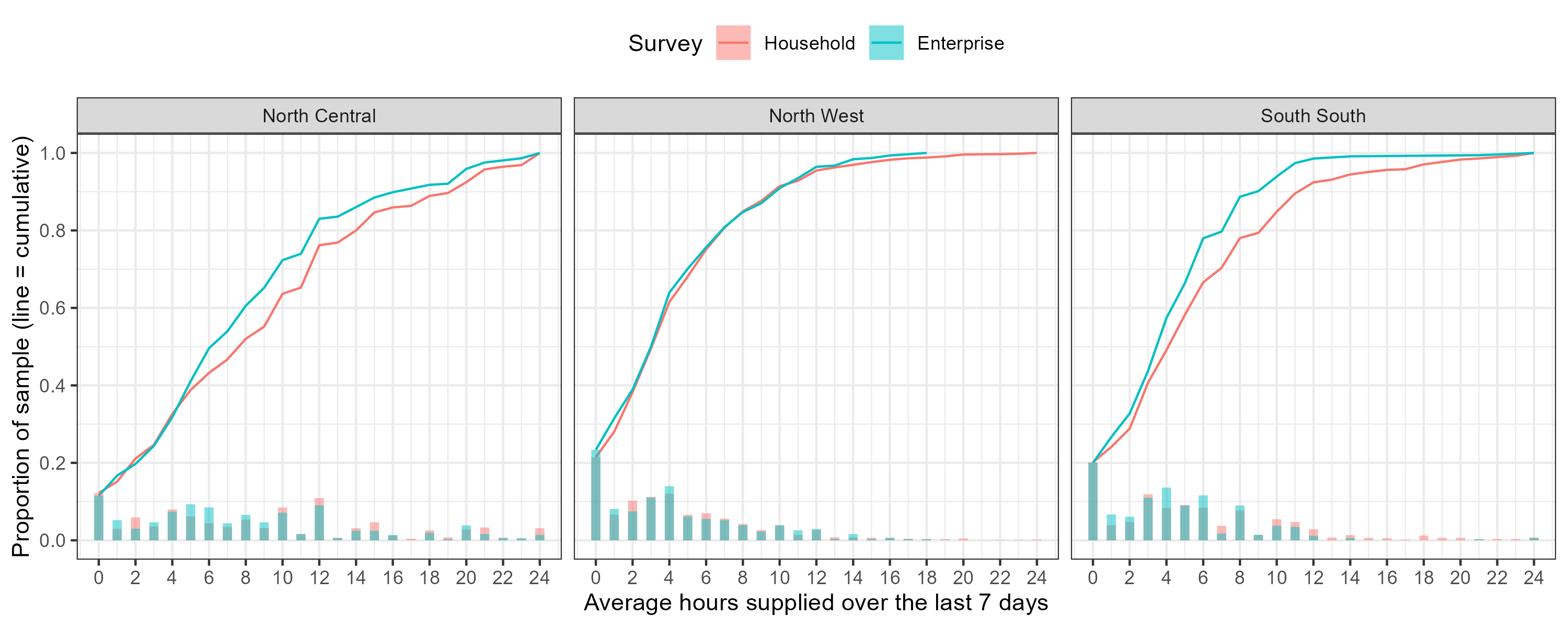}
\caption{\textbf{Distribution of self-reported daily grid electricity supply duration on average over the past week}. The red line shows the cumulative proportion of the household sample that reported between zero and x (the value on the horizontal axis) hours of grid connection supply time per 24 hours. The blue line shows the same for enterprises.}
\label{fig:grid_supplyhrs}
\end{figure}

Table \ref{tab:prisecelec} describes typical primary, secondary and tertiary lighting sources among almost (almost exclusively) grid-connected communities in rural and peri-urban regions of each zone. As noted earlier, we find significant `electricity source stacking' among the sampled households. Dry cell batteries and electric generators are the dominant additional sources of lighting after the national grid, with large differences in the penetration of these sources across the three zones. It is notable that only about half of the households are using the national grid as their primary source of lighting in two of the three zones. Helping to explain this, Figure \ref{fig:grid_supplyhrs}, shows self-reported average daily grid electricity supply duration. Approximately 50\% of both households and enterprises in all zones reported an average less than 8 hours of electricity per day delivered via national grid.

\begin{table}[t!]
\centering
\subtable[Households]{
\scalebox{.67}{

\centering
\begin{tabular}[t]{lrrr}
\toprule
  & Primary & Secondary & Tertiary\\
\midrule
\addlinespace[0.3em]
\multicolumn{4}{l}{\textbf{North Central}}\\
\hspace{1em}National grid connection & 77.2\% & 17.2\% & 4.0\%\\
\hspace{1em}Electric generator & 7.3\% & 26.1\% & 3.4\%\\
\hspace{1em}Rechargeable battery & 2.5\% & 18.5\% & 13.2\%\\
\hspace{1em}Dry cell battery / torch & 12.3\% & 26.6\% & 14.8\%\\
\hspace{1em}Other & 0.7\% & 1.2\% & 6.2\%\\
\addlinespace[0.3em]
\multicolumn{4}{l}{\textbf{North West}}\\
\hspace{1em}National grid connection & 46.8\% & 29.6\% & 8.3\%\\
\hspace{1em}Electric generator & 6.9\% & 8.9\% & 2.2\%\\
\hspace{1em}Rechargeable battery & 7.4\% & 15.2\% & 3.2\%\\
\hspace{1em}Dry cell battery / torch & 32.3\% & 28.0\% & 5.6\%\\
\hspace{1em}Other & 6.4\% & 2.9\% & 4.0\%\\
\addlinespace[0.3em]
\multicolumn{4}{l}{\textbf{South South}}\\
\hspace{1em}National grid connection & 40.9\% & 42.5\% & 10.1\%\\
\hspace{1em}Electric generator & 39.4\% & 23.1\% & 3.3\%\\
\hspace{1em}Rechargeable battery & 3.7\% & 12.8\% & 12.3\%\\
\hspace{1em}Dry cell battery / torch & 10.7\% & 9.2\% & 7.3\%\\
\hspace{1em}Other & 5.3\% & 1.1\% & 4.3\%\\
\bottomrule
\end{tabular}

}}
\subtable[Enterprises]{
\scalebox{.67}{

\centering
\begin{tabular}[t]{lrrr}
\toprule
  & Primary & Secondary & Tertiary\\
\midrule
\addlinespace[0.3em]
\multicolumn{4}{l}{\textbf{North Central}}\\
\hspace{1em}National grid connection & 70.4\% & 22.4\% & 4.8\%\\
\hspace{1em}Electric generator & 17.9\% & 29.9\% & 2.9\%\\
\hspace{1em}Rechargeable battery & 2.7\% & 14.7\% & 8.5\%\\
\hspace{1em}Dry cell battery / torch & 5.6\% & 16.3\% & 5.9\%\\
\hspace{1em}Other & 3.5\% & 2.9\% & 1.3\%\\
\addlinespace[0.3em]
\multicolumn{4}{l}{\textbf{North West}}\\
\hspace{1em}National grid connection & 41.1\% & 35.2\% & 6.5\%\\
\hspace{1em}Electric generator & 34.7\% & 16.9\% & 1.6\%\\
\hspace{1em}Rechargeable battery & 5.1\% & 10.5\% & 3.8\%\\
\hspace{1em}Dry cell battery / torch & 10.5\% & 15.3\% & 7.3\%\\
\hspace{1em}Other & 8.6\% & 3.8\% & 2.2\%\\
\addlinespace[0.3em]
\multicolumn{4}{l}{\textbf{South South}}\\
\hspace{1em}National grid connection & 38.1\% & 48.3\% & 5.9\%\\
\hspace{1em}Electric generator & 52.0\% & 17.9\% & 0.8\%\\
\hspace{1em}Rechargeable battery & 1.9\% & 11.5\% & 8.8\%\\
\hspace{1em}Dry cell battery / torch & 3.7\% & 5.1\% & 4.5\%\\
\hspace{1em}Other & 4.3\% & 0.8\% & 1.1\%\\
\bottomrule
\end{tabular}

}}
\caption{\textbf{Descriptive summary of primary, secondary and tertiary lighting sources at the zonal level.} Households may rely on more than three distinct sources of electricity and lighting in some cases, which we do not show here.}
\label{tab:prisecelec}
\end{table}

\vspace{0.5cm}
\subsection*{Household cookstove stacking}

Cooking with `unclean cooking fuels' is widespread throughout the sampled zones\footnote{Electric cooking, LPG and biogas are classified as "clean cooking" while kerosene and Solid-fuel are among `unclean cooking fuels' \cite{Household_air_pollution2022}.}. The majority of sampled North West households rely primarily on biomass for cooking, with a significant gap between this and LPG. The highest rates of clean cooking are found among South South households with two thirds relying primarily on LPG. Despite grid connection and access to electricity, electric-cooking is not widespread among the sampled households. 

\begin{table}[ht!]
\centering
\scalebox{0.65}{

\centering
\begin{tabular}[t]{lrr}
\toprule
  & Primary & Secondary\\
\midrule
\addlinespace[0.3em]
\multicolumn{3}{l}{\textbf{North Central}}\\
\hspace{1em}Electric & 6.5\% & 12.2\%\\
\hspace{1em}LPG & 38.5\% & 29.8\%\\
\hspace{1em}Biogas & 2.3\% & 1.0\%\\
\hspace{1em}Kerosene & 9.3\% & 29.3\%\\
\hspace{1em}Biomass & 43.5\% & 27.8\%\\
\addlinespace[0.3em]
\multicolumn{3}{l}{\textbf{North West}}\\
\hspace{1em}Electric & 0.5\% & 7.5\%\\
\hspace{1em}LPG & 8.4\% & 21.8\%\\
\hspace{1em}Biogas & 4.8\% & 13.5\%\\
\hspace{1em}Kerosene & 2.5\% & 21.1\%\\
\hspace{1em}Biomass & 83.8\% & 36.1\%\\
\addlinespace[0.3em]
\multicolumn{3}{l}{\textbf{South South}}\\
\hspace{1em}Electric & 1.5\% & 15.2\%\\
\hspace{1em}LPG & 65.0\% & 25.1\%\\
\hspace{1em}Biogas & 1.7\% & 0.5\%\\
\hspace{1em}Kerosene & 18.5\% & 39.8\%\\
\hspace{1em}Biomass & 13.4\% & 19.4\%\\
\bottomrule
\end{tabular}

}
\caption{\textbf{Summary of household cookstove stacking}. Stove use is disaggregated by primary and secondary use which is based on the stove used most with respected to the reported minutes it is operated over a typical day.}
\label{tab:prisecstove}
\end{table}

\subsection*{Appliance ownership}

\begin{figure}[ht]
\centering
\includegraphics[scale = 0.6]{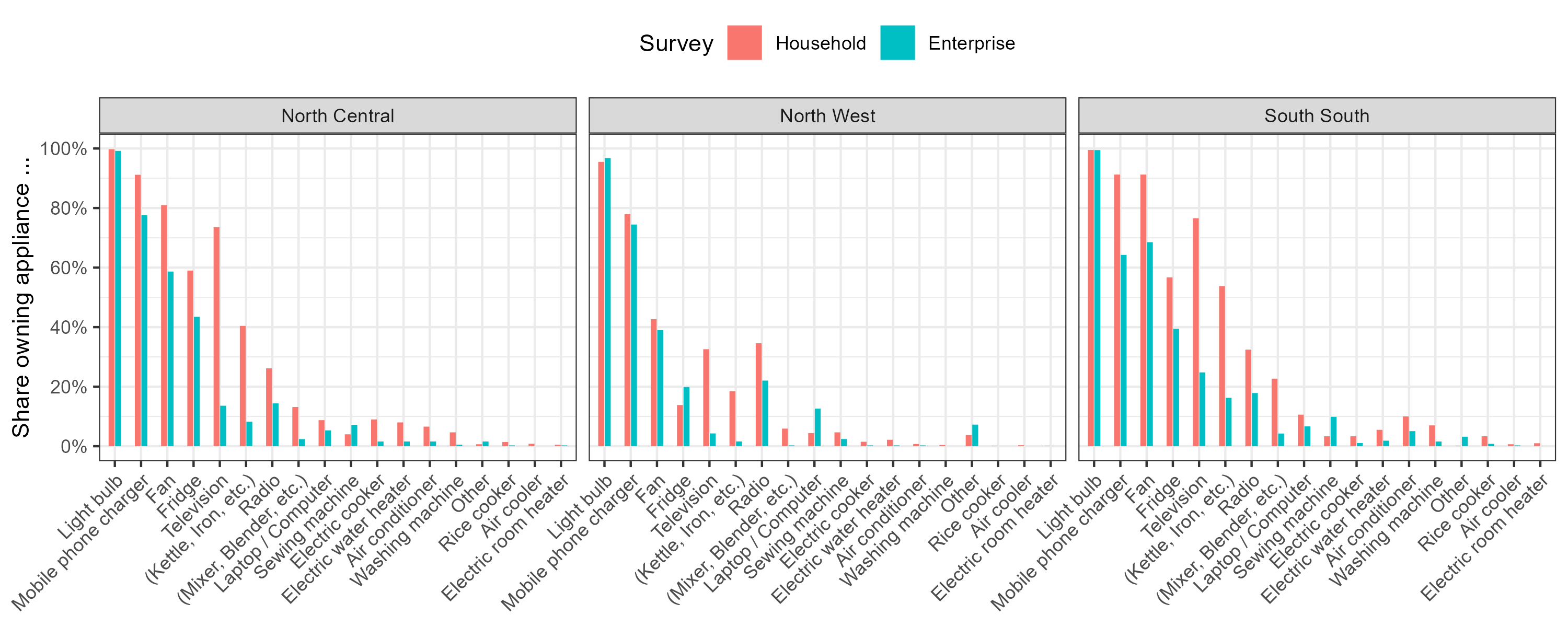}
\caption{\textbf{Appliance ownership rates for each appliance.} Selected aggregate electrical appliance ownership rates for households (red) and enterprises (blue).}
\label{fig:genapp_own}
\end{figure}

Figure \ref{fig:genapp_own} describes the electrical appliance ownership among sampled households and enterprises across the three zones. This figure provides an introduction to the different types of appliances for which ownership and use data were collected. The list of selected appliances queried draws from stakeholder discussions and appliance ownership survey modules in the Multi-Tier Framework for Measuring Energy Access surveys\cite{Bhatia2015}. These ownership patterns reveal the heterogeneity in energy service use among grid-connected households by zone.

\subsection*{Energy service satisfaction}

\begin{figure}[ht!]
\centering
\includegraphics[scale = 0.6]{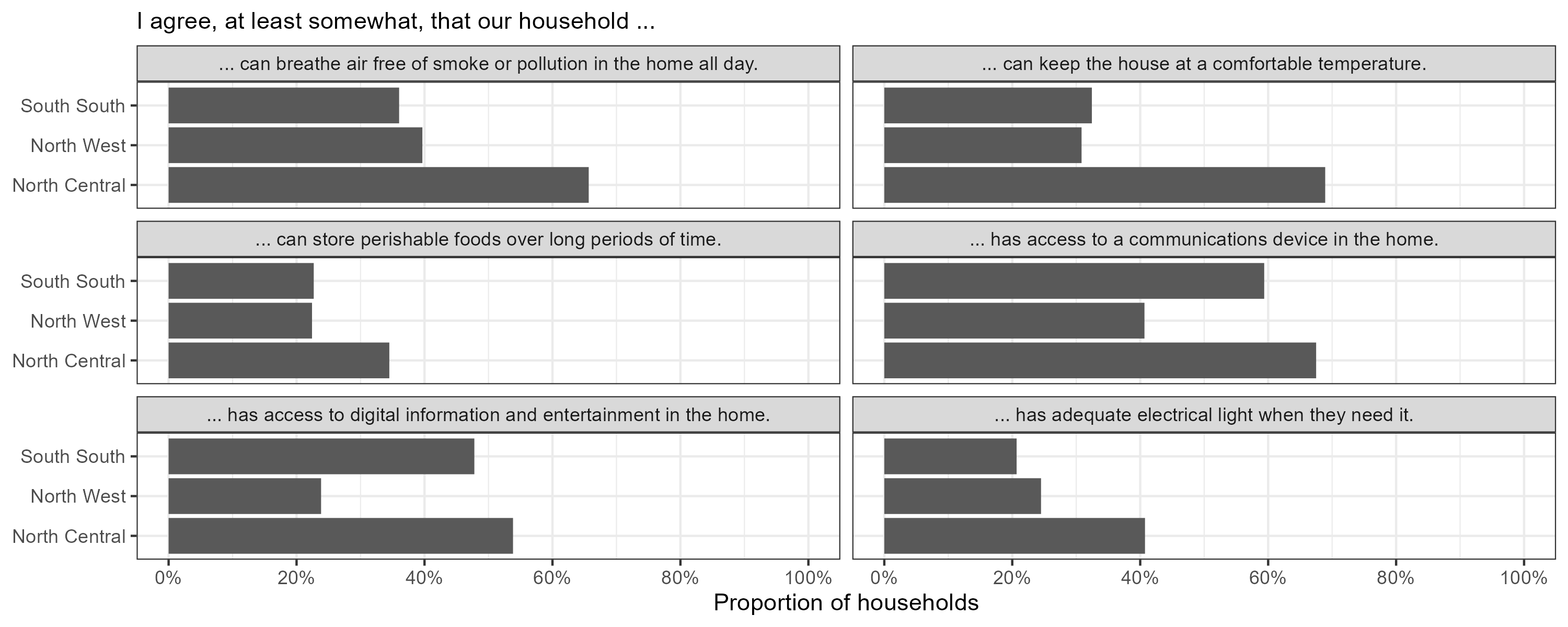}
\caption{\textbf{Reported household energy service satisfaction.} Answers to six questions regarding the electricity service satisfaction.}
\label{fig:enerserve_hh}
\end{figure}

\begin{figure}[ht!]
\centering
\includegraphics[scale = 0.7]{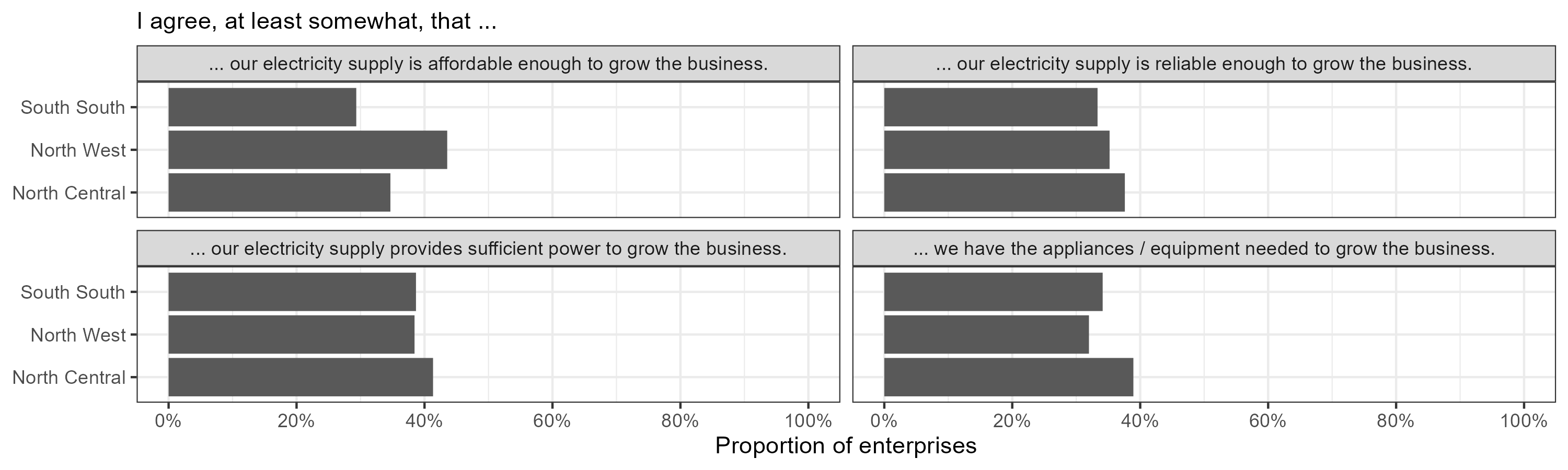}
\caption{\textbf{Reported enterprise energy service satisfaction.} Answers to four questions regarding the electricity service satisfaction.}
\label{fig:enerserve_ent}
\end{figure}

Finally, ten different energy service satisfaction questions are presented in Figures \ref{fig:enerserve_hh} \& \ref{fig:enerserve_ent}. Despite having the largest share of households that own fans and air conditioners, almost two-thirds of South South households believe they are failing to maintain their homes at a comfortable temperature. Regarding their ability to retain perishable food items for extended periods of time and their access to appropriate light when necessary, sampled households across all zones expressed a high level of discontent. North West with the lowest share of phones and televisions in the appliance list, shows the lowest satisfaction with access to communication devices and digital information and entertainment. With regards to enterprises, there is a significant perceived gap between the current energy service levels and what enterprises require for growth. Approximately two-thirds of all surveyed enterprises across all zones were nearly equally unsatisfied with each of the aspects we inquired upon. Heterogeneity across industry, location and size deserves further attention here in subsequent analyses.

\section*{Usage Notes}

We present here a novel survey dataset describing demographic and socioeconomic characteristics, electricity access and supply quality, electrical appliance ownership and usage time, cooking solutions, capabilities, and preferences across households and enterprises living in grid-electrified communities outside the urban core in Nigeria. The dataset is provided as a set of .csv and .xlsx files. Table \ref{tab:usagenotes} describes each file.

\begin{table}[ht!]
\centering
\scalebox{0.65}{
\begin{tabular}{lll}
\hline
Filename & Description & Usage \\ \hline
peoplesun\_hh\_odk\_codebook.xlsx & Household questionnaire codebook & Link to household dataset by variable name \\
peoplesun\_hh\_odk\_choices.xlsx & Household multi- or single-select question labels & Link to codebook by list name \\
peoplesun\_hh\_anon.csv & Anonymised household survey dataset & Apply weights when aggregating zones \\
peoplesun\_hhstoves\_anon.csv & Household stove roster & Link to household dataset by household ID \\
peoplesun\_hhapps\_anon.csv & Household appliance roster & Link to household dataset by household ID \\
peoplesun\_ent\_odk\_codebook.xlsx & Enterprise questionnaire codebook & Link to enterprise dataset by variable name \\
peoplesun\_ent\_odk\_choices.xlsx & Enterprise multi- or single-select question labels & Link to codebook by list name \\
peoplesun\_ent\_anon.csv & Anonymised enterprise survey dataset & Apply weights when aggregating zones \\
peoplesun\_entstoves\_anon.csv & Enterprise stove roster & Link to enterprise dataset by enterprise ID \\
peoplesun\_entapps\_anon.csv & Enterprise appliance roster & Link to enterprise dataset by enterprise ID \\
peoplesun\_entequips\_anon.csv & Enterprise equipment roster & Link to enterprise dataset by enterprise ID \\
eaidgeokey\_anon.csv & Basic geospatial data per enumeration area & Link to datasets by enumeration area ID \\ \hline
\end{tabular}}
\caption{\textbf{Overview of files included in the presented dataset.}}
\label{tab:usagenotes}
\end{table}

This dataset can inform decision makers, project developers and researchers about the current energy supply situation and model future energy use towards securing a decent living standard for all in rural and peri-urban areas of Nigeria. Our preliminary assessment points to three key areas of further research using this data. First, we propose \textit{modelling appliance ownership likelihoods, electricity consumption levels and energy service needs in un-electrified regions}. Surveyed use of appliances among grid-connected communities can serve as a proxy to estimate future electricity demand for similar households which are currently not grid connected. We find the appliances used are mainly light bulbs, phone charger, fans, television and kettles / irons and fridges (with a higher variance across the zones). Appliance ownership is combined with detailed data describing patterns of acquisition and use, aiding modelling efforts. Second, we propose \textit{identifying supply-side and demand-side solutions to address high usage of diesel generators}. The collected data provides evidence of very unreliable grid supply among rural and peri-urban grid-electrified communities. In all zones, approximately 50\% of households and enterprises reported receiving less than 8 hours of electricity per day from the national grid. A significant proportion of enterprises and households cope with this situation by using diesel generators as primary or secondary lighting source. Thirdly, we propose \textit{exploring broader issues of multi-dimensional energy access, access to decent living standards and climate vulnerability}. The data indicates that households in all zones generally expressed high levels of dissatisfaction regarding their to access basic energy services such as keeping oneself at a comfortable temperature, though heterogeneity across the zones was evident. With regard to productive uses of electricity, roughly 60\% of all surveyed enterprises are dissatisfied with energy supply affordability, reliability and adequacy, alongside their access to necessary appliances.

A common theme across these three research areas is the importance of increasing the central grid supply quality and in providing cost-competitive decentralized alternatives to improve living conditions and business perspectives in Nigeria. Although our research focusses on energy access, we understand that achieving SDG7 - providing sustainable energy for all - is not the silver bullet for increasing living conditions and business perspectives on its own. A myriad of accompanying factors such as access to other basic services including education, sanitation and health, access to capital and broader issues of gender equality must be addressed. Nevertheless, SDG7 is a key enabling achievement as many other SDGs depend upon sustainable energy access. With this in mind, it is clear that achieving improving energy access in Nigeria is both about new connections and about increasing the supply quality in weak grid areas. The existing grid network needs to be strengthened and decentralised renewables will play a role in both peri-urban and rural regions. Furthermore, in order to satisfy energy service needs in the context of a changing global climate, demand-side policies supporting household acquisition of appliances necessary for a decent living standard must be explored. A cross-cutting analysis combining expertise from institutions responsible for centralised network planning and those responsible for decentralised electrification efforts would be ideal to strategically plan these efforts. We hope that the data we present can be useful in providing evidence for such an analysis.

\section*{Code availability}
As we are presenting collected survey data, no custom code was used or is necessary to generate or work with this data. Complete codebooks describing the questionnaires are available here: \url{
https://dataverse.harvard.edu/dataset.xhtml?persistentId=doi:10.7910/DVN/GTNEJD}. 

\bibliography{ref}

\section*{Acknowledgements}

This research work is part of the three-year research and development project "People Power: Optimizing off-grid electricity supply systems in Nigeria" (PeopleSuN). PeopleSuN is a project funded by the German Federal Ministry of Education and Research (BMBF) within the funding initiative "Client II - International Partnerships for Sustainable Innovations" (FKZ 03SF0606A). We thank eHealth Africa for their impeccable work in collecting the data and their cooperation. We would also like to thank Tomiwa Aileru, Clara Bruegge, Oriyomi Amusa and Ulrike Lich who supported us in the preparation of the survey instruments and processing of the data.

\section*{Author contributions statement}

S.P., C.B, and P.B. conducted stakeholder discussions to define the data gap (for details, see https://www.peoplesun.org/). S.P. developed the sampling and data collection methods. eHealthAfrica implemented the data collection. S.P. conducted the analysis presented here. All authors contributed in drafting the accompanying text.

\section*{Competing interests}

The authors declare no competing interests.

\section*{Figures \& Tables}

\end{document}